\documentclass{article}

\usepackage[final]{neurips_2020_ml4ps}
\usepackage[utf8]{inputenc} 
\usepackage[T1]{fontenc}    
\usepackage{hyperref}       
\usepackage{url}            
\usepackage{booktabs}       
\usepackage{amsfonts}       
\usepackage{nicefrac}       
\usepackage{microtype}      
\usepackage{graphicx}
\usepackage{dcolumn}
 \usepackage{bm}
\title{Accelerating MCMC algorithms through Bayesian Deep Networks }
\usepackage{ulem}
\usepackage{xcolor}

\usepackage{fontawesome}
\newcommand{\github}{\href{https://github.com/JavierOrjuela/BayesianNeuralNets_CMB}{\faGithub}}
\author{%
  H\'ector Javier Hort\'ua \\
  Grupo  de  Gravitaci\'on  y  Cosmolog\'ia,\\
  Observatorio Astron\'omico Nacional\\
  Universidad  Nacional  de  Colombia,\\
  \texttt{hjhortuao@unal.edu.co} 
  \And
   Riccardo Volpi\\
   Romanian  Institute  of  Science  and  Technology (RIST) \\
   Cluj-Napoca,  Romania \\
   \texttt{volpi@rist.ro} \\
   \And
   Dimitri Marinelli \\
   FinNet project and \\ Munich Re Markets, Munich\\
   \texttt{dm@financial-networks.eu} \\
   \And
   Luigi Malagò \\
   Romanian  Institute  of  Science  and  Technology (RIST) \\
   Cluj-Napoca,  Romania \\
   \texttt{malago@rist.ro} \\
}

\begin{document}

\maketitle

\begin{abstract}
Markov Chain Monte Carlo (MCMC) algorithms are commonly used for their versatility in sampling from complicated probability distributions. However, as the dimension of the distribution gets larger, the computational costsfor a satisfactory exploration of the sampling space become challenging.  Adaptive MCMC methods employing a choice of proposal distribution  can address this issue speeding up the convergence. In this paper we show an alternative way of performing adaptive MCMC, by using the outcome of Bayesian Neural Networks as the initial proposal for the Markov Chain. This combined approach increases the acceptance rate in the Metropolis-Hasting algorithm and accelerate the convergence of the MCMC while reaching the same final accuracy. Finally, we demonstrate the main advantages of this  approach  by constraining the cosmological parameters directly from  Cosmic  Microwave  Background  maps.\github
\end{abstract}

\section{\label{sec:MCMC} Introduction}
Cosmological observations have significantly increased  in the last decade allowing us to obtain a better description of the Universe. This task has been also achieved thanks to  Bayesian inference methods allowing to derive constraints on the parameters of  cosmological models from those observations~\cite{verde2007practical}. Bayesian inference offers a way to learn the prediction task from data through the posterior distribution $p(\theta|d)\sim p(d|\theta)p(\theta)$;  being $\theta$  a set  of  unknown  parameters  of interest,  $d$ the data associated with a measurement, $p(\theta)$ the  prior distribution  that quantifies what we know about $\theta$ before observing any data, and  $p(d|\theta)$  is the likelihood  function.
Computing the true posterior is generally intractable, and approximation methods must be implemented in order to perform Bayesian inference in practice. Two main techniques  for  this  purpose  are Variational Inference and Markov Chain Monte Carlo (MCMC)~\cite{NIPS2011_4329,doi:10.1063/1.1699114,regier2018approximate,jain2018variational}. The former  method although  computationally faster, requires  the  approximation  of  the  true  posterior, while  the latter  has become one of the most popular  methods for cosmological parameter estimation due  to  its  advantage of being non-parametric and asymptotically exact.
Classical  MCMC methods  draw samples sequentially according to a probabilistic algorithm that allows to scale linearly with the dimension of the parameter space~\cite{verde2007practical}. However if the complexity of the model increases from  the presence of "slow" parameters, nuisance parameters, foregrounds or parameter correlations,  the sampling will exhibit a high numerical  cost~\cite{PhysRevD.87.103529}. Additionally, 
it is generally difficult to determine a convenient initial state for the system and an accurate criterion to determine the convergence of the Markov Chain.  These practical issues compel MCMC practitioners to resort  convergence diagnostic tools which demand to run MCMC for a very long time to obtain  good solutions~\cite{10.1111/j.1365-2966.2004.08464.x}.
In this work we show a preliminary approach to accelerate the  convergence of MCMC by including in the bottom of it, the approximate  distribution outcome of the deep neural network as a  proposal for the Markov Chain. We show the  advantages of this approach  in  constraining the cosmological parameters directly from Cosmic Microwave Background  (CMB) maps.
\section{Dataset and Network}
\label{sec:DSBNN}
Following~\citet{hortua2019parameters}, we use 50.000 images related to the CMB  maps  projected  in $20\times20\, $deg$^2$ patches in the sky for training the Bayesian Neural Network. These images have a dimensions (256,256,3), where the last channel stands for the Temperature (channel=0) and Polarization (channel=1,2), and each image corresponds to a specific set value of the cosmological parameters: dark matter density $\omega_{cdm}$, spectrum  amplitude $A_s$, baryon density $\omega_b$, and the dark energy density parameter $\Omega_\Lambda$. The BNN was implemented in TensorFlow-Probability, and the same  version of the VGG architecture along with the presence of Flipout as it was shown in~\cite{hortua2019parameters}  was used in this paper. Finally, we used the calibration method introduced in~\cite{hortua2020reliable} where $\alpha$-divergence with $\alpha=1$ has been included at the top of the BNN. Results of the conditional distributions for the predicted parameters  by our BNN vs standard MCMC methods
are displayed in Fig.~\ref{fig:mcmc}.  
\section{Method}
MCMC algorithms are commonly used for sampling from complicated distributions.  As the dimension of the distribution gets larger, the computational costs
for a satisfactory exploration of the sampling space become challenging.
Adaptive  MCMC methods such as the choice of proposal distributions in the  Metropolis-Hastings algorithms  are designed to address this issue  speeding  up the convergence. However, a suitable class of distribution is almost never known in advance and  the search for improved proposal distributions is often done manually, through trial and error, which can be difficult especially in high dimensions. The method shown in this paper can be seen as  a novel way to perform adaptive MCMC in which the output distribution of the BNNs serves as a proposal distribution for the MCMC. As it was shown in the previous work of ~\citet{hortua2019parameters}, multi-channel BNNs, are able to break degeneracies among parameters and provide reliable results close to the desired conditional posterior. This distribution can be used as potential proposals to  significantly  improve  the  performance  during  parameter inference.  MCMC experiments were run in the cobaya software~\cite{torrado2020cobaya}, with the likelihood given by~\cite{verde2007practical}
\begin{eqnarray}
  -\mathcal{L} &\sim& \sum_l(2l+1)\bigg[\ln\bigg(\frac{C_l^{BB}}{\hat{C}_l^{BB}}\bigg(\frac{C_l^{TT}C_l^{EE}-(C_l^{TE})^2}{\hat{C}_l^{TT}\hat{C}_l^{EE}-(\hat{C}_l^{TE})^2}\bigg)\bigg)\nonumber\\
  &+& \frac{\hat{C}_l^{BB}}{C_l^{BB}}+\frac{\hat{C}_l^{TT}C_l^{EE}+C_l^{TT}\hat{C}_l^{EE} -2\hat{C}_l^{TE}C_l^{TE}}{C_l^{TT}C_l^{EE}-(C_l^{TE})^2}\bigg],
\end{eqnarray}
where $\hat{C}_l$ is the power spectrum of the CMB patches  obtained with Lens-Tools~\cite{2016A&C....17...73P}, and $C_l$  the theoretical model. Cobaya accepts the cosmological parameters as input, compute $C_l$ via CLASS~\cite{Blas_2011} and when the Markov chains have enough points to provide reasonable samples from the  posterior distributions, the simulation stops and it  return  the chains. We run two MCMC experiments taking into account the power spectrum of the CMB maps.  In the first MCMC experiments we used the full sky map, while in the second one, we computed the power spectra for CMB patches and used them as an input in cobaya package.
\section{Results}
\begin{figure}[h!]
\begin{center}
\includegraphics[width=0.68\textwidth]{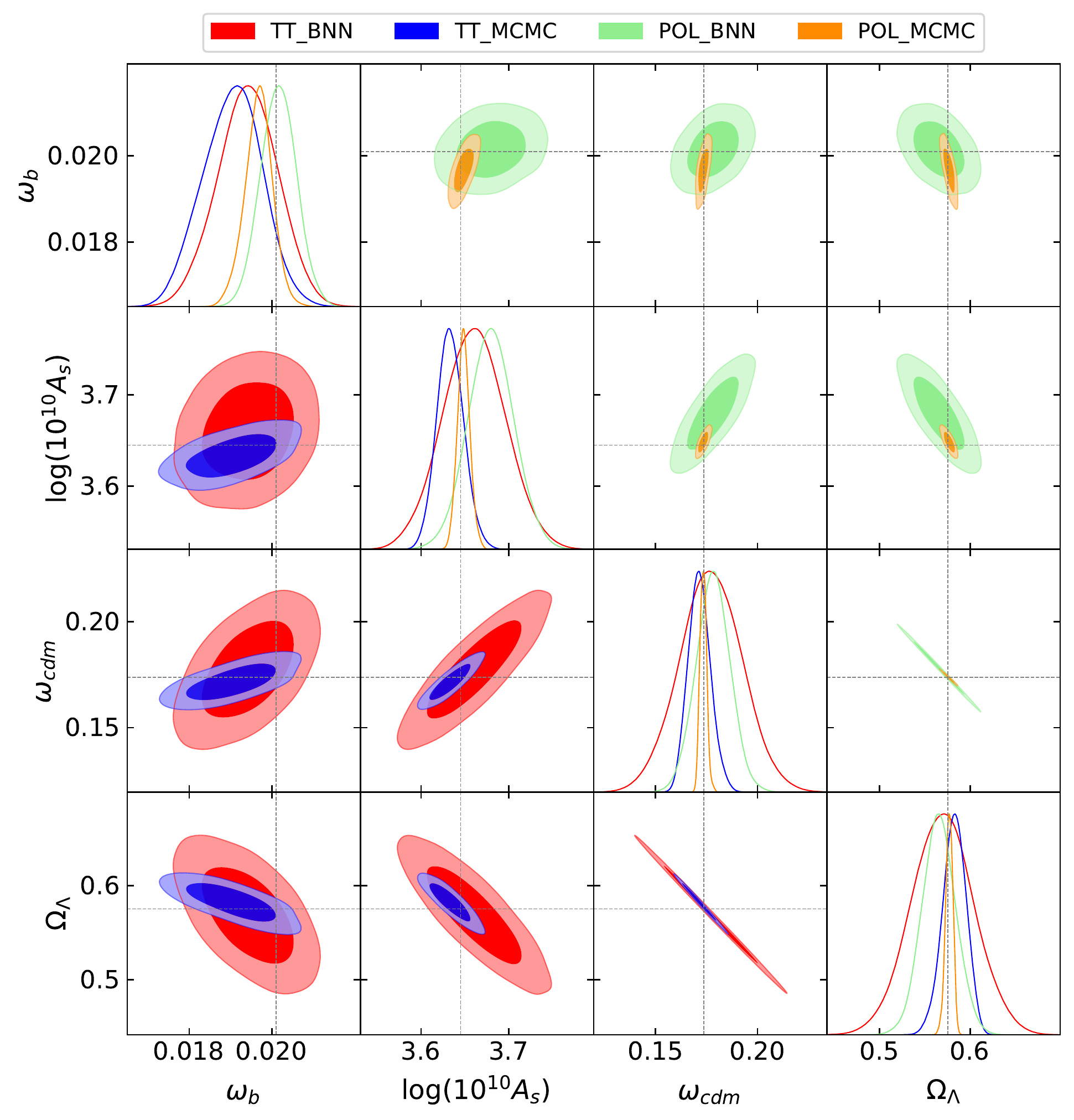}
\end{center}
\caption{\it  Marginalized  parameter  constraints  obtained from temperature maps (TT) and combined temperature  with polarization (POL) using MCMC and the best BNN model.  The black line stand for the real value: $\omega_b=0.0201$, $\log(10^{10}A_s)=3.6450$ and  $\omega_{cdm}=0.1736$ taken from the test dataset. } \label{fig:mcmc}
\end{figure}

Results of the conditional distributions for the predicted parameters are displayed in Fig.~\ref{fig:mcmc} where we compared the MCMC results with the calibrated BNNs (on the CMB patches described in Section~\ref{sec:DSBNN}).

\begin{table}[h!]
  \caption{\it Statistics and Parameters $95\%$ intervals for the minimal base-$\Lambda$CDM model from our synthetic CMB dataset using non-informative  priori (MCMC) and a  precomputed covariance matrix from VI (covarBNN). The last column reports the metrics using the Full CMB map. The bold values in the last column correspond to the implementation of a  proposal posterior distribution from VI. Although the full sky gives the smallest credible region, MCMC is 10000 times slower than VI. The real value considered is the same as specified in Fig.~\ref{fig:mcmc}} \label{table:mcmctimes}
  \centering
  \scalebox{0.87}{
\begin{tabular}{|l||l|l|l|l|l|l|}
\hline
\multicolumn{6}{|c|}{Statistics  for various MCMC sampling configurations}                                                                                                                    \\ \hline
\multicolumn{1}{|l||}{\textbf{Metrics}} & \multicolumn{2}{l|}{\textbf{Temperature map}}& \multicolumn{3}{l|}{\textbf{Temperature+Polarization map}}  \\
\cline{2-6}\hline\rule{0pt}{10pt}
 &\multicolumn{1}{l|}{\textbf{MCMC}}& \multicolumn{1}{l|}{\textbf{covarBNN}} & \multicolumn{1}{l|}{\textbf{MCMC}} & \multicolumn{1}{l|}{\textbf{covarBNN}}& \multicolumn{1}{l|}{\textbf{Full-sky}} \\
\hline\rule{0pt}{10pt}
$ \omega_{b}$                                           & $0.0190^{+0.0013}_{-0.0013}$ &  $0.0190^{+0.0012}_{-0.0012}$ & $0.01967^{+0.00066}_{-0.00066}$ &  $0.01968^{+0.00064}_{-0.00064}$            &       $ 0.02009^{+0.00010}_{-0.00010}$     \\ \hline
\rule{0pt}{10pt}$\ln(10^{10}A_s)$                       &$3.633^{+0.031}_{-0.031}    $    &$3.633^{+0.031}_{-0.030}   $ & $3.648^{+0.015}_{-0.015}   $   &               $3.648^{+0.015}_{-0.016}     $&     $3.6449^{+0.0027}_{-0.0027}  $     \\ \hline \rule{0pt}{10pt}
$\omega_{cdm}$                                           & $0.171^{+0.011}_{-0.011}   $   &$0.170^{+0.011}_{-0.011}   $ & $0.1734^{+0.0031}_{-0.0032}   $   &          $0.1734^{+0.0031}_{-0.0031}   $   &   $0.1736^{+0.0009}_{-0.0009}$    \\ \hline\rule{0pt}{10pt}
$\Omega_{\Lambda}$                                        &  $0.583^{+0.025}_{-0.025}   $  &  $0.583^{+0.024}_{-0.025}   $ &  $0.5769^{+0.0079}_{-0.0080}   $   &           $0.5769^{+0.0079}_{-0.0079}   $&$ 0.5793^{+0.0019}_{-0.0019}$       \\ \hline\rule{0pt}{10pt}
Runtime                                                  &$4.02$hr &$1.56$hr &$4.40$hr &$3.14$hr &$4.52$hr//${\bf 3.15}$hr \\ \hline\rule{0pt}{10pt}
Acc. rate                                                  &$0.19$ &$0.23$ &$0.14$ &$0.25$ &$0.18$//${\bf 0.23}$\\ \hline\rule{0pt}{10pt}
$R-1$                                                  &$0.0093$ &$0.0098$ &$0.0051$ &$0.0084$ & $0.0091$//${\bf 0.0090}$\\ \hline\rule{0pt}{10pt}
$(R-1)_{95\%CL}$                                                  &$0.0827$ &$0.0764$ &$0.0944$ &$0.0642$ & $0.0940$//${\bf 0.0800}$\\ \hline
 \end{tabular}}
\end{table}
\begin{figure}[h!]
\begin{center}
\includegraphics[width=0.88\textwidth]{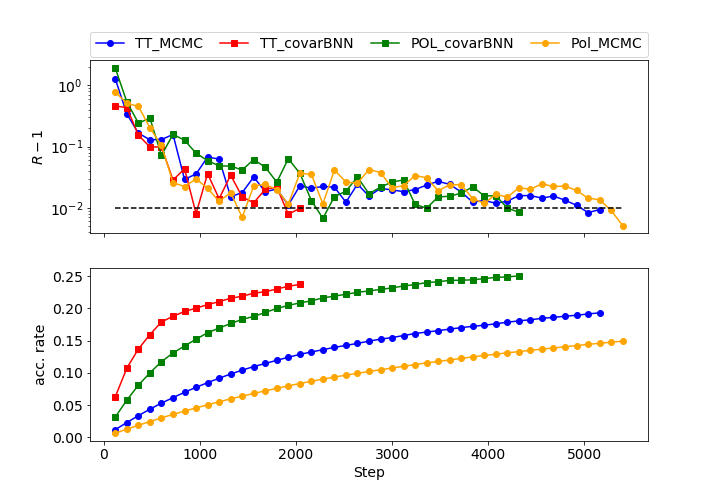}
\end{center}
\caption{\it  Graphical representation of convergence in MCMC using non-informative  priori (MCMC) and a  precomputed covariance matrix from VI (covarBNN).  (Top)  Gelman-Rubin values with respect to  the acceptance step. (Bottom) Acceptance rate with respect the acceptance step. Notice that using a  proposal  covariance matrix  takes shorter time for chains to converge. } \label{fig:mcmc_step}
\end{figure}
We  observe that  MCMC provides  tighter and more accurate constraints. However, the trained Neural Network can generate 8000 samples in approximately ten seconds which it turns out to be 10000 times faster than MCMC for this dataset~\footnote{We run all single  MCMC experiments in  a CPU  Intel Core i7-3840QM with clock speed of 2.80GHz, while the BNN was trained in a GPU: GeForce GTX 1080 Ti.}. Runtime and metrics for convergence in  MCMC are shown in Table~\ref{table:mcmctimes}. As  expected, the polarization combined with temperature data  shifts the values obtained from temperature alone and  enhances the  accuracy in all parameters (columns 1 and 3). Furthermore, a well-converged chain  is also observed via the  Gelman-Rubin $R-1$ parameter and its standard deviation  at  $95\%$ confidence level interval $(R-1)_{95\%CL}$ (the smaller the better).
The qualitative correlations among parameters as obtained from BNNs are mostly analogous to the MCMC ones (Fig.~\ref{fig:mcmc}), showing that a multi-channel BNN is able to handle the complexities involved in this kind of analysis and additionally to use the polarization information to break cosmological degeneracies.
Nonetheless, although 10000 times slower, the MCMC it is able to better quantify the uncertainty. This is especially true when using the power spectrum of the full map, and the intervals are an order of magnitude more accurate than those computed by VI (rightmost column of Table~\ref{table:mcmctimes}).  On the other hand, we can also combine MCMC and VI leveraging the advantages of both methods. Such topic has attracted a lot of attention in the recent literature~\cite{10.5555/3045118.3045248,thin2020metflow}. A straightforward approach to speed up MCMC algorithms consists in  using the covariance matrix constructed from the chains of the trained Neural Network as proposal for the distribution of the MCMC. In fact, it is known that a good estimate of the  covariance matrix for the parameters
increases the  acceptance rate leading to significantly faster convergence~\cite{PhysRevD.87.103529}.
In Table~\ref{table:mcmctimes}, we compare the runtime for the MCMC with and without a precomputed covariance obtained from BNN. As we can see from the table, proposal covariances from BNNs (covarBNN) speed up convergence in MCMC reducing the  computational time for all datasets (Temperature, Polarization and full sky maps). In Fig.~\ref{fig:mcmc_step} we report MCMC convergence diagnostic quantities such as $R-1$ and the acceptance rate per iteration. The stopping  rule implemented in cobaya  ensures that the Gelman-Rubin $R-1$ value and its standard deviation  at  $95\%$ confidence level interval $(R-1)_{95\%CL}$ computed  from  different chains (four in our case), satisfy the convergence criterion $R-1<0.01$ twice in a row, and  $(R-1)_{95\%CL}<0.2$ respectively to stop the run,~\cite{torrado2020cobaya}. For the Temperature signal alone, the Markov chains achieve a steady state in about 2000 steps working with the covarBNN proposal while it usually takes more than 5000 steps instead with the vanilla MCMC. This behavior can also be explained by observing the acceptance rate in Fig.~\ref{fig:mcmc_step} (bottom), the red curve (TTcovarBNN) quickly approaches a considerably high acceptance rate, eventually converging at around 0.23 (which is a standard value for  which we expect to have a  decent  acceptance rate~\cite{roberts1997}). An analogous trend can be seen for the polarization case.

\section{Discussion}
In this paper we show that MCMC algorithms  excel  at  quantifying  uncertainty with respect to  BNNs models, although the latter  is  about  10000  times  faster  at inference.  Given these properties, we showed  an approach in which the covariance matrix efficiently estimated from the BNNs samples,  significantly  enhance  the  acceptance rate in MCMC yielding faster convergence. A limitation of this method is  the use of partial  CMB maps that prevent the access to large scales correlations, leading to large uncertainties and possibly introducing a bias in the prediction of the cosmological parameters sensible to such scales. It would be interesting (and more of a fair comparison) to compare MCMC for a full sky with respect to spherical neural architectures~\cite{2019arXiv190204083K,Perraudin:2018rbt,cohen2018spherical}  which can extract large scale signals correlations, thus  determining if Deep Learning methods can achieve a similar level of precision as compared with MCMC.
As a future work, we also expect to assess the performance of this method with respect to other MCMC modifications such as a 
Hamiltonian Monte Carlo.

\section*{Broader Impact}
Combining the speed of Bayesian Neural Networks with the accuracy of MCMC algorithms results in potential Bayesian inference methods for upcoming cosmological observation. Additionally, the network built in this work allows to adaptively extract complicated correlations when performing inference without assuming a priori  summary statistics such as power spectrum or higher order spectra (such as bispectrum, trispectrum or others).
This work presents an example for how machine learning could be used in the physics  community to improve classical inference methods. 
\section*{Acknowledgments}
{\bf Software  used} Argo library (https://github.com/rist-ro/argo) for training the Bayesian Neural Network; cobaya~\cite{torrado2020cobaya} was used for MCMC sampling; Healpy~\cite{Zonca2019} \& LensTools~\cite{2016A&C....17...73P} for computing the power spectra from images; CLASS~\cite{Blas_2011} for obtaining the theoretical power spectrum. The  data generator script and
MCMC chains are available at \github. \\

R.~Volpi, and L.~Malag\`o are supported by the DeepRiemann project, co-funded by the European Regional Development Fund and the Romanian Government through the Competitiveness Operational Programme 2014-2020, Action 1.1.4, project ID P\_37\_714, contract no. 136/27.09.2016. 
H.J.~Hort\'ua, acknowledge the RIST institute where they were employed when this projected started and the initial support from the DeepRiemann project.

\medskip
\small
\bibliography{neurips_2020}




\end{document}